\documentclass[aps,prl,showpacs,twocolumn,groupedaddress]{revtex4-1}
\usepackage{epsfig}
\usepackage{dcolumn}
\usepackage{epstopdf}
\usepackage{graphicx}
\usepackage{ulem}
\usepackage{color}

\begin{document}
\title{Ferromagnetism and orbital order in a topological ferroelectric}
\author{Marco Scarrozza}
\author{Alessio Filippetti}
\author{Vincenzo Fiorentini}
\affiliation{CNR-IOM, UOS Cagliari,  and Department of  Physics, University of Cagliari, Cittadella Universitaria, I-09042 Monserrato (CA), Italy}

\date{\today}

\begin{abstract}
We explore   via density functional calculations the magnetic doping of a topological ferroelectric  as an unconventional route to multiferroicity.  Vanadium doping of the layered perovskite La$_{2}$Ti$_{2}$O$_{7}$ largely preserves electric polarization and produces robust  ferromagnetic order, hence proper multiferroicity. The marked tendency of dopants to cluster   into chains results in an insulating character at generic doping. Ferromagnetism  stems from  the  symmetry breaking of the multi-orbital V system via an unusual ``antiferro''-orbital order, and from the host's   low-symmetry  layered structure.
\end{abstract}
\maketitle

Multiferroic materials, where ferroelectricity  and magnetism  coexist, have been studied intensively in recent years \cite{Picozzi2009}
due to their fascinating physical properties and their potential for  application (e.g. multi-state or electrically switchable  magnetic memories) \cite{Eerenstein2006}. 
Despite the effort, single-phase bulk multiferroics at room temperature are still unavailable; most are antiferromagnetic (AF) and not, as desired, ferromagnetic (FM); and,  generally, they do not meet \cite{Martin2008} the integration and functional requirements of device applications \cite{Bea2008,Binek2005}.   A parallel line of research  \cite{Martin2008,Ramesh2007}  on multi-phase systems (interfaces, heterostructures, nanostructures) revealed unconventional ferroelectricity mechanisms in artificially  layered perovskites  (e.g. SrTiO$_{3}$/PbTiO$_{3}$ superlattices \cite{Bousquet2008}). Recently, it was suggested \cite{Benedek2011,Lopez2011} that magnetism and ferroelectricity will be more cooperative  in lower-symmetry, naturally nano-structured, layered  systems. In one such system, the layered perovskite oxide La$_{2}$Ti$_{2}$O$_{7}$ (LTO), ferroelectricity is not displacive  but rather results from dipoles generated by antiferrodistortive oxygen-octahedra rotations, which fail to cancel out due to the layered structure,  resulting in a net macroscopic polarization \cite{Lopez2011}. Because ferroelectricity follows directly from its  layered structure, LTO was labeled  \cite{Lopez2011}  a topological ferroelectric. LTO  has a  T$_{{\rm C}}$ of 1770 K, a sizable  polarization $P_{c}$=5 ${\mu}$C/cm$^{2}$ \cite{explto} along the $c$ crystal axis (schematic in Fig.2), and non-critical permittivity ${\varepsilon}$$\simeq$50 which would be an asset  for  integration in ferroelectric random-access memory elements \cite{Atuchin2009}. 

Here we  explore theoretically a route to multiferroicity  based on doping LTO with magnetically active ions substituting for Ti (nominally a 4+ 3$d^{0}$ ion). By an extensive screening  of transition-metal dopants, we found vanadium (V, nominally 4+ and 3$d^{1}$ in LTO) to  yield robust ferromagnetic order while preserving polarization and a small gap, thus giving rise to proper multiferroicity. We find that V align in chains along the $a$ axis, orthogonal to the polarization $c$ axis (with  magnetic moments orthogonal to the chains themselves),  resulting in insulating character at generic doping. Intriguingly, ferromagnetic coupling is linked to the layered structure. The latter ``pre-wires'' a two-orbital-per-site system, whose degeneracy is broken in favor of ferromagnetic superexchange by a peculiar orbital ordering, as foreshadowed in earlier theoretical work \cite{Kugel1982}.  We preliminarily investigate the magnetoelectric response to (a subset of) ionic vibrations, finding it to be almost absent. 

We perform ab initio calculations  within spin-density functional theory (DFT) in the generalized gradient approximation (GGA) \cite{Perdew1992} 
as implemented in the VASP code \cite{Kresse1996}, with  projector-augmented waves \cite{Blochl1994} for La 5$spd$6$s$, Ti and V 3$pd$4$s$, and O 2$sp$ states,  plane-wave cutoff  330 eV, and  a 4$\times$2$\times$5 grid in the primitive cell's Brillouin zone, folded for larger supercells. Magnetic calculations for magnetoelectricity and anisotropy are non-collinear and include spin-orbit coupling (see below). GGA seems a reliable  approach for LTO, which is a band insulator with O $p$ valence and Ti $d$ conduction \cite{Lopez2011}; even the V electrons (see below) sit into conduction-derived states, whose the localization is considerable but not extreme.

\begin{figure}[ht]
\includegraphics[width=1\linewidth]{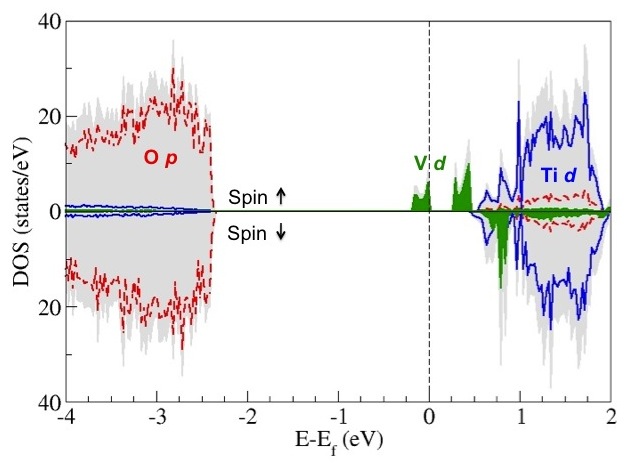}
\caption{\label{fig1}Total density of states (DOS) of LTO:V (gray shade) and its site projections for O (dashed red line), Ti (solid blue line), and V (green shade).}
\end{figure}

LTO is the $n$=4 member of the A$_{n}$B$_{n}$O$_{3n+2}$ family, and
consists of four perovskite-like units (distorted TiO$_{6}$ octahedra) stacked along the [011] cartesian direction and separated periodically by the insertion of an extra O layer \cite{Lichtenberg2001}. We optimize cell and internal parameters (force threshold 20 meV/\AA) in the experimentally-stable monoclinic $P2_{1}$ structure. 
Our calculated lattice parameters $a$=7.80 \AA, $b$=13.22 \AA, $c$=5.58 \AA, and ${\gamma}$=98.52$^{\circ}$ are in excellent agreement with experiment \cite{Gasperin1975}.   The  energetics and electronic structure of V-doped LTO is obtained  collating data from calculations on 44-atom and 88-atom cells at 12.5\% average V concentration. Our conclusions do not really depend critically on this figure, which in essence only amounts to approximate the isolated-impurity limit with a single V at no less than about 8 \AA\, from  its periodic image. 
The  isolated V's  spin-polarized electron yields a magnetic moment of 1.0 ${\mu_{B}}$. Its spin-resolved and atom-projected density of states (DOS)  shown in Fig.\ref{fig1} is typical of all FM configurations of V discussed below. The valence band is mostly of O $p$ character, and the conduction band is mostly of O-hybridized Ti $d$ character. Near the Fermi level,  V induces two spin-majority  peaks, one  carrying  the extra electron and the other empty, across a gap of 0.3 eV. Ti states start 0.5 eV, and minority  V $d$ states  at  $\sim$0.85 eV above the edge of the occupied V band.

\begin{figure}
\includegraphics[width=1\linewidth]{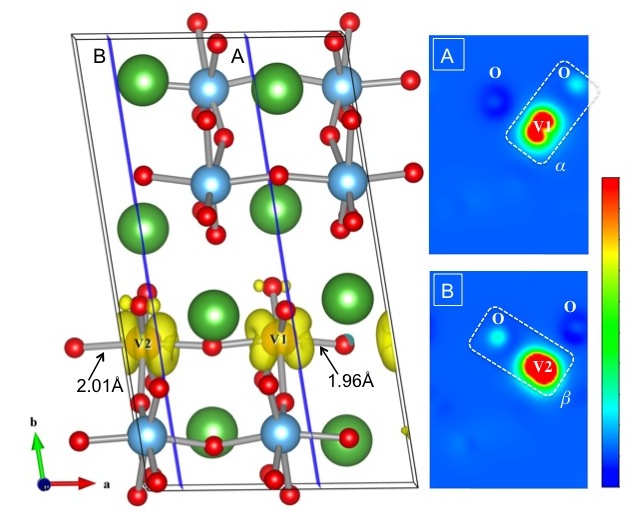}
\caption{\label{fig2} On the right: Isosurface plot of the spin density for the LTO:V in V-chain configuration; on the left: 2D cuts of the spin density along the two A and B planes, 
normal to the $a$ axis, in proximity of the V1 and V2 atoms, respectively.}
\end{figure}

To  estimate the magnetic  coupling between  V's, we  sampled  a variety of
 structural (V pairs on the relevant non symmetry-equivalent cation B sites) and magnetic (collinear FM and AF) configurations. Structurally, we find  a  tendency to chain ordering of V along the $a$ direction (see Fig.\ref{fig2}). Chains are favored by 100 meV per V pair over the first competing configuration, and 180 meV per V pair referred to isolated impurities. Magnetically, FM order is very stable within the  V   chains, with FM couplings $J_{a}$$\simeq$80 meV and 20 meV along the $a$ direction, respectively, for chains in proximity of the extra O layer or inside the perovskite block. Both types of chain (or fragment thereof, see below) will appear in similar concentrations as their energies differ by only about 5 meV/dopant. The interchain coupling in the $b$,$c$ plane is also FM and small, $J_{b,c}$$\simeq$3 meV at 12\% V.

For a ballpark estimate of the critical  temperature we use the  Ising model  with axial ani\-so\-tro\-py \cite{yuri}, whose critical temperature T$_{\rm C}$ is determined by the ratio $R$=$J_{{b,c}}$/$J_{a}$ of in-plane  to on-axis coupling. The  model mirrors closely our system of weakly coupled chains with strong internal coupling along one axis.  We estimate the FM T$_{\rm C}$ for both our two nearly-degenerate ground states with  distinct $J_{a}$'s, since   magnetic couplings are  not additive. For $J_{a}$$\simeq$80 meV, $R$=0.04  and correspondingly T$_{\rm C}$$\sim$0.93$J_{a}$=860 K; for $J_{a}$$\simeq$20 meV, $R$=0.15  and T$_{\rm C}$$\sim$1.55$J_{a}$=360 K, i.e. T$_{\rm C}$ is around room temperature in the  worst-case estimate.
 
We now check that  the material remains ferroelectric upon doping.  The polarization  calculated via the Berry phase technique \cite{resta} for LTO is  $P$$^{0}_{c}$=5.2 $\mu$C/cm$^{2}$ along the $c$ axis, similarly to experiment and previous calculations \cite{explto,Bruyer2010}. In V-doped LTO,  the distortions generating ferroelectricity are fully conserved, aside from minor quantitative differences. 
We find a slightly enhanced and reduced $P_{c}$ for the energetically-favored chain  configurations, respectively, inside the blocks and near the additional O layers. 
Since polarization is extensive, we average the values for these   nearly-degenerate configurations obtaining $P_c$=4.7 $\mu$C/cm$^{2}$, i.e. only 10\% lower than undoped LTO at 12.5\% V concentration. From $P$ values at 12\% and 25\% V, we infer that $P_{c}$[$x$]$\simeq$$P^{0}_{c}$(1--$x$) with $x$ \% V. 
Interestingly, V doping produces polarization components along $a$ and $b$,  4.5 $\mu$C/cm$^{2}$ and 0.7 $\mu$C/m$^{2}$ respectively at 12\% V, comparable with that along $c$. We will address the consequences of this elsewhere.

So far we saw that V-doped LTO is insulating, polarized, and  ferromagnetic, and therefore properly multiferroic. To show that this applies at generic doping, we evaluate energetics, magnetism, and insulating character of V  chain fragments in LTO comparing three configurations with 4 cations lined up along $a$: $i$) -V-V-V-V-, $ii$) -V-V-Ti-Ti-, and $iii$)  -V-Ti-V-Ti-. The full chain $i$) is bound  by  100 meV compared to two separated V dimers (chain fragments) as in $ii$). In turn, the V dimer fragment $ii$) is 130 meV lower in energy than the ``broken'' dimer $iii$). Energy thus drives V to cluster in homogeneous chains; these will be  in fact finite chain fragments accounting for configurational entropy (we will elaborate on this elsewhere.) All chain and chain-fragment configurations are insulating, and $i$) and $ii$) have similar  FM coupling  along $a$ as measured by the FM-AF energy difference; these are clear indications  that the system will be insulating and FM at a generic concentration  due to the V tendency to chain clustering. 

\begin{figure}
\includegraphics[width=1\linewidth]{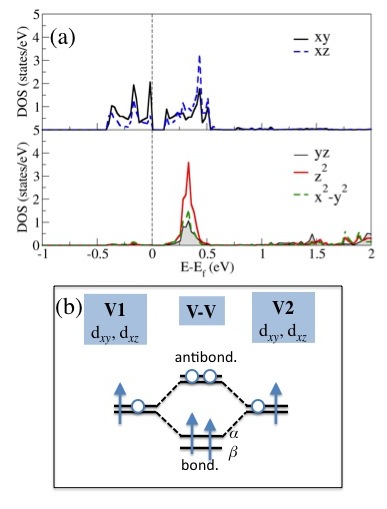}
\caption{\label{fig3}(a) Projected orbital resolved DOS relative to a V atom in LTO:V with 2V/cell, the V-chain configuration.
(b) Schematic energy levels diagram for interacting V-V pair in LTO:V.}
\end{figure}

We now analyze the strong ferromagnetic coupling. V-doped LTO is at first glance a  degenerate multi-orbital system. However, this degeneracy is resolved so as to favor FM order, and the mechanism is, to a significant extent, built-in into the host structure. {\it First}, only two orbitals per V site are involved: the V 3$d$ DOS of a full V chain in  Fig.\ref{fig3}a shows $d_{xy}$, $d_{xz}$ states lower in energy and involved in V-V bonding and  magnetism, while  $d_{yz}$ follows the  $e_{g}$ states to higher energy. Indeed, in the  2$'$ point group of LTO the octahedral-symmetry $t_{2g}$  and $e_{g}$    representations  split into 1-fold $a_{1}$ and $b_{1}$. $d_{yz}$ and $e_g$ functions belongs to  $a_{1}$, whereas $d_{xy}$ and $d_{xz}$ belong to $b_{1}$. Physically, one notices that the periodic alignment of octahedra along their axes is interrupted by the layered structure, except along the $a$ axis on which $d_{xy}$ and $d_{xz}$ overlap significantly.
{\it Second}, the bonding-antibonding combination (Fig.\ref{fig3}) of $d_{xy}$ and $d_{xz}$  along the chain generates two narrow  bands separated by a small gap (see e.g. \cite{Mahadevan2004} for a similar case). {\it Third}, because of  the  large energy gain due to Hund on-site exchange,  FM order  will clearly be favored over AF (whose much smaller gain is of order the superexchange energy), provided that, {\it fourth}, the two-orbital bonding combination is non-degenerate, and hence  different on each V site, so as to host two spin-parallel electrons. Indeed, Fig.\ref{fig3}a clearly shows the different weight of the two orbitals in the  bonding state on a specific site; this alternation  is accompanied by modulated bond lengths and by differences in V magnetic moments, as shown below. The resulting  FM spin order and  ``antiferro''-orbital order for the  orbital filling agree with early predictions \cite{Kugel1982} as well as with Goodenough-Kanamori rules \cite{Goodenough1963}. The chain-like ordering of V, we note, is actually  relevant even in some binary vanadates, as discussed e.g. in \cite{vanadates}.
  
The computed spin density shown in Fig.\ref{fig2} further  confirms the above argument. 
It is localized on  V atoms and has the fingerprint shape of  a $t_{2g}$-like combination.  The 2D cuts of the spin density along planes (A and B in the figure) normal to the $a$ axis  
and close to the V atoms,  clearly show ``antiferro''-orbital order. The spin density lobes on the two V sites are  distinctly slanted towards alternating O atoms in the upper layer, one roughly normal to the other. The V chain has resolved the degeneracy by alternating $d_{xy}$-$d_{xz}$ orbital combinations ${\alpha}$ and $\beta$, respectively with 65/35\%  and 54/46\% $d_{xy}$/$d_{xz}$ content.
As hinted-to above, concomitantly with spin and orbital ordering, the V-chain is also characterized by dimerized bond lengths and moments  (see Fig.\ref{fig2}):  pairs of short (1.95 and 1.97 \AA) and long (2.00 and 2.02 \AA) O-V-O bonds  alternate along the chain direction, and the magnetic moments of short-bond and long-bond V are also alternating (0.75 and 1.03 $\mu_{B}$). As seen in Fig.\ref{fig2}, state $\beta$ is centered on the long-bond large-moment V$_{2}$ atom and $\alpha$ on V$_{1}$.

 To further  validate   the   link between  V dimerization and orbital ordering, we compare  V-doped LTO in $P$2$_{1}$ symmetry with the parent high-symmetry structure with space group $P2_{1}/m$.
This structure has no  antiferro-distortive  rotations and only one inequivalent V site along each   $a$ chain, with V-V distances 3.91 \AA\, and  180$^{\circ}$ O-V-O angle. Thus, it cannot remove the orbital degeneracy, and the bonding-antibonding splitting does not materialize. The system ends up being metallic and ferromagnetic, with identical spin-density  on  all V atoms. The structural dimerization is evidently an integral part of the explanation of the insulating nature and the spin and orbital ordering in V-doped LTO, as its absence causes  orbital degeneracy and metallicity. In this very light, and in keeping with our previous symmetry considerations, we note that the O-Ti-O bond lengths along $a$ are already modulated in LTO (1.97, 2.03, 1.98, and 2.02 \AA). The structural inequivalencies along the $a$ axis are thus at least partly pre-wired in the  host structure's  low symmetry, supporting the general idea of a search for  multiferroicity opportunities in unconventional, low symmetry materials.

Since multiferroicity and magnetoelectricity (the coupling of, e.g., electric field to magnetization) are often associated,  we study preliminarily  the lattice component of the magnetoelectric tensor, expressed by \'I\~niguez  \cite{iniguez} as the product of polarization and magnetization changes in response to polar-mode distortion divided by the mode force constant.  We need magnetic non-collinear spin-orbit calculations with  full spin relaxation and a  stricter convergence limit than usual ($<$\,0.01 $\mu$eV). We consider small displacements away from the  ferroelectric $P$2$_{1}$ structure along the composite IR mode (mainly rotations around $a$) connecting it to the   paraelectric $P$2$_{1}$/$m$ structure. To numerical accuracy, all magnetization components  are insensitive to this composite displacement.  A rationale  is provided by magnetic anisotropy calculations, whereby we constrain the spin direction to  map out the energy vs spin orientation. The V spins  preferentially point  normal to the chain direction--that is, $a$ is  the hard  axis, with a substantial anisotropy energy  180 $\mu$eV/ spin with respect to the $b$,$c$ plane. The $b$,$c$ plane, in turn, hosts  many spin-configurational minima within less than 5 $\mu$eV/spin, i.e. we  all directions in the $b$,$c$ plane are equally easy: thus, spins ``freewheel'' around the $a$ axis irrespective of the rotations  (we dub this  the ``ball-bearing'' effect) and are insensitive to octahedra rotations around $a$. Interestingly, V spins are  always  approximately parallel, and they point near the [011] direction in the $b$,$c$ plane in the lowest-energy configuration.
In conclusion, for the composite ferro-para IR mode  the lattice  magnetoelectric tensor is zero. This may not be the whole story, however, as we did not consider all possible IR modes; some of those we neglected may couple to the $a$ and $b$ polarization components appearing in V-doped LTO. Also, we ignored  the electronic component, which might still be appreciable. 

Finally, we investigate the effect of oxygen vacancies (which are donor defects)  on  magnetism, motivated by our finding for Sc-doped LTO that the vacancy formation energy drops to almost zero, destroying ferromagnetism. In equilibrium oxygen-rich conditions (chemical potential $\mu_{\rm O}$=$\mu_{\rm O_{2}}/2$), the vacancy concentration is negligible at typical growth temperatures, because the lowest (as function of configuration) formation energy of the vacancy  in the neutral state in LTO is 5.4 eV.  In V-doped LTO, the  concentration is still negligible, as the formation energy is 4.6 eV. Consistently with  the lower formation energy, the vacancy   binds  to V pairs within the $a$ chain, turning the magnetic ordering  from FM to AF. This    suggests the possibility to  spatially  structure the magnetic state by controlling the local O content in layer-by-layer growth techniques.

In summary, the V-doped layered ferroelectric LTO  shows several unique properties of basic as well as  applicative (e.g. data storage) interest. It is a ferromagnetic multiferroic with high temperature ferroelectricity and  ferromagnetism at any doping, due to V clustering in longish, homogeneous chains.  Polarization is significant, and  decreases mildly as function of V concentration (at least at small doping).  Magnetoelectric coupling seems negligible at the present approximation level.  Control of native defects could  turn FM  order into AF, enabling the engineering of  resistivity and magnetic phase changes via defect injection and migration. 

Work supported in part by  IIT via   Seed project NEWDFESCM,  by MIUR via project 2DEG-FOXI, and  by Fondazione Banco di Sardegna. Computing resources: CASPUR, CINECA,  Cybersar.


\begin{thebibliography}{99}

\bibitem{Picozzi2009}
S. Picozzi and C. Ederer, J.  Phys.: Condens. Matt. {\bf 21}, 303201 (2009);
C.	Ederer	and	N.	A.	Spaldin,  Curr. Op.  Sol. State  Mat. Sci. {\bf 9}, 128 (2005);
D. Khomskii, Physics {\bf 2}, 20 (2009).

\bibitem{Eerenstein2006}%
 W. Eerenstein, N. D. Mathur,	and J. F. Scott, Nature {\bf 442}, 759 (2006).
 
\bibitem{Martin2008}
L. W. Martin, S. P. Crane, Y.-H. Chu, M. B. Holcomb,	M.	Gajek,	M.	Huijben,	C.-H.	Yang, N. Balke, and R. Ramesh,
J.  Phys.: Cond. Matt. {\bf 20}, 434220 (2008).

\bibitem{Bea2008}%
H. B\'ea, M. Gajek, M. Bibes, and A. Barth\'el\'emy, J.  Phys.: Cond. Matt. {\bf 20}, 434221 (2008).

\bibitem{Binek2005}%
C. Binek and B. Doudin, J.  Phys.: Cond. Matt. {\bf 17}, L39 (2005).

 

\bibitem{Ramesh2007}
R. Ramesh, and N. A. Spaldin, Nature Materials {\bf 6}, 21 (2007).

\bibitem{Bousquet2008}
E. Bousquet, M. Dawber, N. Stucki, C. Lichtensteiger, P. Hermet,
S. Gariglio, J.-M. Triscone, and P. Ghosez, Nature {\bf 452}, 732
(2008).

\bibitem{Benedek2011}%
N.	A.	Benedek	and	C.	J.	Fennie, Phys. Rev. Lett. {\bf 106}, 107204 (2011).


\bibitem{Lopez2011}%
J. L\'opez-P\'erez and J. \'I\~niguez,
Phys. Rev. B  {\bf 84}, 075121 (2011).

\bibitem{explto}
S. Nanamatsu, M. Kimura, K. Doi, S. Matsushita, and N. Yamada,
Ferroelectrics {\bf 8}, 511 (1974).

\bibitem{Atuchin2009}
V. V. Atuchin, T. A. Gavrilova, J.-C. Grivel, and V. G. Kesler, J. Phys. D: Appl. Phys. {\bf 42}, 035305 (2009).
\bibitem{Kugel1982}%
K. I. Kugel and D. I. Khomskii, 
Sov. Phys. Usp. {\bf 25}, 231 (1982).

\bibitem{Perdew1992}%
J. P. Perdew, J. A. Chevary, S. H. Vosko, K. A. Jackson, M. R. Pederson, D. J. Singh, and C. Fiolhais, Phys. Rev. B {\bf 46}, 6671 (1992).

\bibitem{Kresse1996}%
G. Kresse and J. Furthm\"uller, Phys. Rev. B {\bf 54}, 11169 (1996);
G. Kresse and D. Joubert,  Phys. Rev. B {\bf 59}, 1758 (1999).

\bibitem{Blochl1994}%
P. E. Bl\"ochl, Phys. Rev. B {\bf 50}, 17953 (1994).

\bibitem{Lichtenberg2001}
F. Lichtenberg, A. Herrnberger, K. Wiedenmann, and J. Mannhart, Prog.  Sol. St. Chem. {\bf 29}, 1 (2001); F. Lichtenberg, A. Herrnberger, and K. Wiedenmann, Prog.  Sol. St. Chem. {\bf 36}, 253 (2008).

\bibitem{Gasperin1975}
M. Gasperin, Acta Cryst.  B {\bf 31}, 2129 (1975). 

\bibitem{yuri}
M. Yurishchev, arXiv:cond-mat/0312555.

\bibitem{resta}
R. Resta, J. Phys.: Condens. Matter {\bf 12}, R107 (2000).


\bibitem{Bruyer2010}%
E. Bruyer and A. Sayede, 
J. Appl. Phys. {\bf 108}, 053705 (2010)



\bibitem{Mahadevan2004}
P. Mahadevan, A. Zunger,	and D. D. Sarma, Phys. Rev. Lett. {\bf 93}, 177201 (2004).

 
\bibitem{Goodenough1963}
J. B. Goodenough, {\it Magnetism and the Chemical Bond} (Wiley-Interscience, New York 1963); J. B. Goodenough, Scholarpedia {\bf 3}, 7382 (2008).

\bibitem{vanadates}
V. Eyert, Phys. Rev. Lett. {\bf 107}, 016401 (2011); Ann. Phys. (Leipzig) {\bf 11}, 650 (2002).
\bibitem{iniguez}
J. \'I\~niguez, Phys Rev. Lett. {\bf 101}, 117201 (2008).

\end{thebibliography}
\end{document}